\documentclass[12pt]{article}

\usepackage{amsmath}
\usepackage{graphicx}
\usepackage[font=small,margin=10pt,labelfont=bf]{caption}
\usepackage{subcaption}
\usepackage{array}                    
\usepackage[linkbordercolor={0.28 0.28 0.43}]{hyperref}
\usepackage{multirow}
\usepackage[table]{xcolor}
\usepackage{authblk} 

\begin{document}

\title{Modeling the Frictional Driving of a Gyroscope Casing by a Spinning Rotor}

\author[1]{Vedat Tanr{\i}verdi\thanks{Email: \href{mailto:tanriverdivedat@googlemail.com}{tanriverdivedat@googlemail.com} \\ ORCID: \href{https://orcid.org/0000-0003-0225-2183}{0000-0003-0225-2183}}}
\affil[1]{Ankara Science University, 06570 Ankara, Turkey}

\author[2]{Arda Erbasan\thanks{Email: \href{mailto:arda.erbasan@metu.edu.tr}{arda.erbasan@metu.edu.tr} \\ ORCID: \href{https://orcid.org/0000-0001-6630-8241}{0000-0001-6630-8241}}}
\affil[2]{Department of Physics, Middle East Technical University, 06800 Ankara, Turkey}

\date{\today}

\maketitle

\begin{abstract}
The rotation of the casing and rotor of a gyroscope is studied by considering frictional effects.
Friction causes the casing to rotate, and over time, air dissipation and friction at the touchpoint gradually stop this rotation.
Several models for air friction, friction between the rotor and casing, and friction at the touchpoint are analyzed.
Fit results demonstrate that while some of these models can describe the primary motion, certain effects require further study to yield more precise results.
These findings can aid in developing improved models for the rotation of satellites.
\end{abstract}

\section{Introduction}

Friction can affect the motion of a gyroscope and a heavy symmetric top in different ways.
For example, without friction, we cannot explain the rise of a heavy symmetric top \cite{Jellett, Fokker1941, Tanriverdi_rise}.
In addition, friction is an important factor in the rotation of space satellites, as some utilize mechanical gyroscopes for attitude control \cite{starin}.

Various aspects of friction regarding the motion of a heavy symmetric top have been previously studied.
The study of damped free oscillations in a gyroscopic system considering linear (viscous) friction is one of the pioneering works in this area \cite{ParksMaunder}.
High-spin gyroscopic systems have also been studied, assuming a linear damping term \cite{SethnaBalachandra}.
Dissipation proportional to the $n$th power in a gyroscopic system has been investigated alongside linear dissipation \cite{GeWu}.
The spinning of a rigid body under an external torque is studied in \cite{LonguskiTsiotras}.

Gyroscopes are used as Control Moment Gyros (CMGs) in space satellites and telescopes for attitude control.
A history of CMG development, found in \cite{Sorokin}, provides insight into the developmental process, although it is predominantly Soviet-based.
Friction is a crucial factor in the study of CMGs.
Dahl provided one of the first friction models for sliding and rolling friction by primarily considering stiction and a constant friction term \cite{Dahl}.
Later, de Wit et al. improved the friction model by developing the LuGre model, which includes a viscous term \cite{deWit, astrom}.
The LuGre model has since been widely adopted \cite{deWit2, pan2020, hill, gaiandwang, sancakbayraktaroglu, iwasaki}.
A broader survey of friction models is given by Olsson et al. \cite{Olsson}.

Another aspect of this problem is the use of ball bearings in CMGs.
Friction in these ball bearings exhibits complex properties \cite{wu_he}.
While the Dahl model provides a foundation for capturing pre-sliding hysteresis \cite{Dahl}, the LuGre model extends this framework to accurately reproduce low-speed phenomena, such as stick-slip properties and the Stribeck effect \cite{deWit, astrom}.
Ball bearing friction can depend on a fractional power of the angular velocity \cite{wu_he} or a fractional power of the normal force \cite{Olsson}.
However, this property is not included in these models.
Although these models are highly successful at low angular velocities, they have shortcomings at high angular velocities, where temperature can also affect friction \cite{wolf_iskender}.

In this work, unlike the previous studies, we present an investigation of the case where the friction causes the rotation of the casing.
Similar methods are used to study ball bearing friction but it is not used to study the motion of a casing \cite{wu_he}.
This study focuses on the kinetic phase of the motion of the gyroscope, and hence the initial motion is excluded from the scope. 
Therefore, the static friction and the Stribeck effect are not involved in the motion.
Possible ball bearing friction effects and other potential models in which the air dissipation is also included are examined through two experimental cases.
Besides these, a joint fit is used.
Furthermore, energy considerations are applied to test the different models.

In section 2, the model developed to explain the motion is described.
The experiment is presented in section 3.
Then, different possible models are tested in section 4.
In section 5, two different models are used to determine the friction parameters.
In section 6, an energy approach is applied, and finally, a summary and conclusion are presented in section 7.

\section{Model}

The rotation of a casing resulting from the frictional force between a gyroscope's rotor and its casing can provide a deeper understanding of frictional effects.
In the experiment, the rotor is set into rotation.
Its initial rotation provides the necessary energy to rotate the casing, and as time passes, the rotation slows down due to dissipation.
Videos of the experiment can be viewed using the links provided in the appendix.

In this case, the rotation is one-dimensional, and a single angle is sufficient to explain the rotation: $\psi$.
The rotation angles of the rotor and casing are represented by $\psi_r$ and $\psi_c$, respectively.

There are several dissipative forces present, which can be modeled with the following equations:
\begin{eqnarray}
    I_r \ddot \psi_r &=& -\tau_{cr} - a_1 \dot \psi_r -a_2 \dot \psi_r^2, \nonumber \\
    I_c \ddot \psi_c &=&  \tau_{cr} -b_1-b_2 \dot \psi_c, \label{dieqns}
\end{eqnarray}
where $I_r$ is the moment of inertia of the rotor, and $I_c$ is that of the casing.
$\tau_{cr}$ represents the frictional effects between the rotor and the casing, which slow down the rotor and drive the rotation of the casing. 
$a_1$ and $a_2$ are considered as results of the air dissipation for the rotor.
$b_1$ and $b_2$ represent dissipative effects for the casing, originating from air dissipation and friction at the touchpoint between the casing and the ground.

It is possible to add another term for the air dissipation of the casing depending on $\dot \psi_c^2$; however, tests have shown that the fit quality changes only very slightly with this term.
Hence, it was removed from the model.

For the frictional effects between the casing and the rotor, the following model is employed by considering different possible interactions:
\begin{equation}
    \tau_{cr}=c +d |\dot \psi_r -\dot \psi_c|+e (\dot \psi_r -\dot \psi_c)^2+f |\dot \psi_r -\dot \psi_c|^g \label{frictiont}
\end{equation}
where $c$ is related to the constant interaction, $d$ to the linear interaction, and $e$ to the quadratic interaction. 
The parameters $f$ and $g$ are included since the friction between the casing and rotor mainly occurs due to the rotation of the ball bearing.

\section{Experimental setup and data}

\begin{figure}[h!]
\centering
    \includegraphics[width=0.5\textwidth]{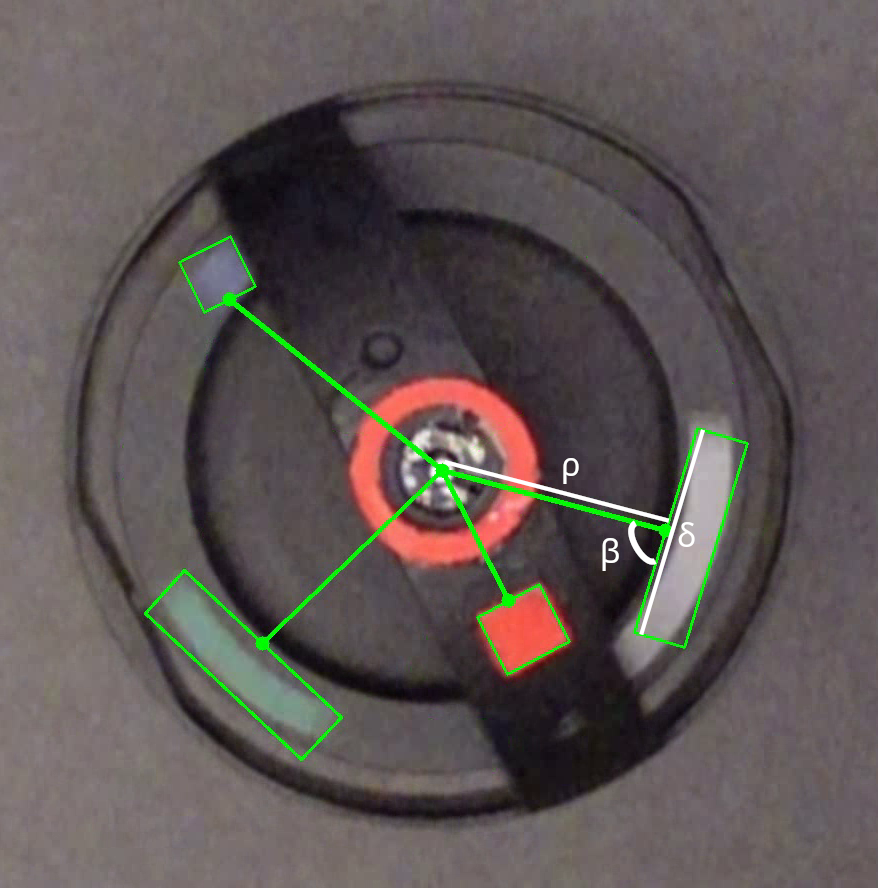}
  \caption{Measurements of the gyroscope's angular positions.
    The (red) circle at the center is used to find the position of the center. The (red) rectangular piece is used to measure the angular positions of the casing.
    Three rectangular pieces (blue, white, and green) are used to measure the angular positions of the rotor. Because of the fast rotation, they appear smeared.
        The corners of the rectangular pieces nearest to the center are used to determine the angular positions. 
    For each rectangular piece, the base length $\delta$, the radius $\rho$, and the angle between the radius and base $\beta$ are calculated.}
  \label{fig:gyroscope_p}
\end{figure}

For the experiment, a commercial gyroscope is used, and colored paper pieces are utilized to measure the angular positions (see figure \ref{fig:gyroscope_p}).  
Two experimental results are studied in this work.
For case 1, a coated cardstock is used as the contact surface.
For case 2, a clear PVC sheet is used.
The change in the contact surface may affect parameters $b_1$ and $b_2$.

\begin{figure}[h!]
\centering
    \includegraphics[width=0.5\textwidth]{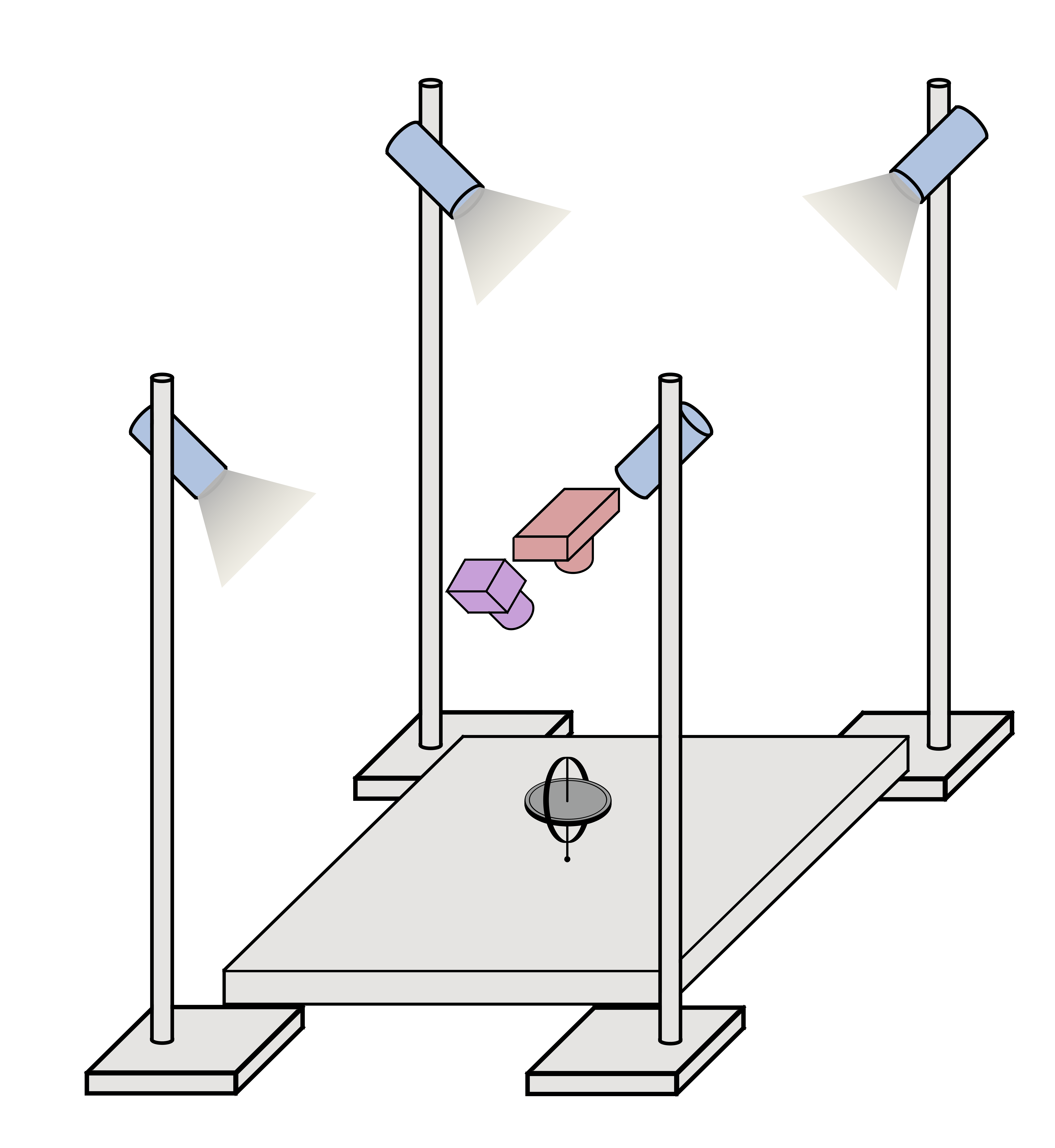}
  \caption{ Experimental setup. 
	The gyroscope is at the center. Four lights (blue) are used to iluminate it. Sony ZV-1 camera (red) is at the top of the gyroscope, and ELP Web camera (purple) is next to it.
    }
  \label{fig:exp_setup}
\end{figure}

\begin{figure}[h!]
\centering
  \begin{subfigure}[b]{0.45\textwidth}
    \includegraphics[width=\textwidth]{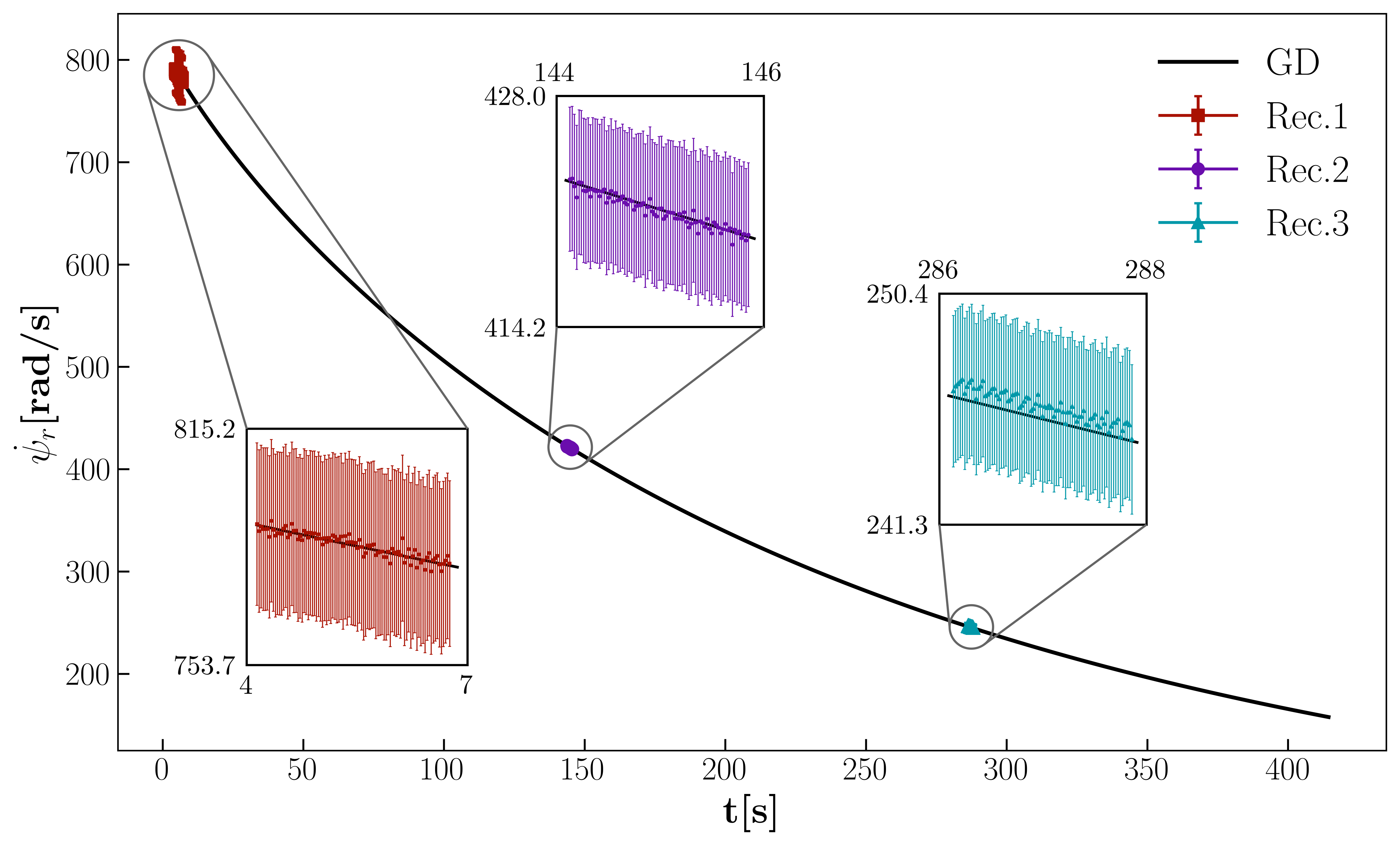}
    \caption{Case 1}
    \label{fig:first_gd}
  \end{subfigure}
  \hfill
  \begin{subfigure}[b]{0.45\textwidth}
    \includegraphics[width=\textwidth]{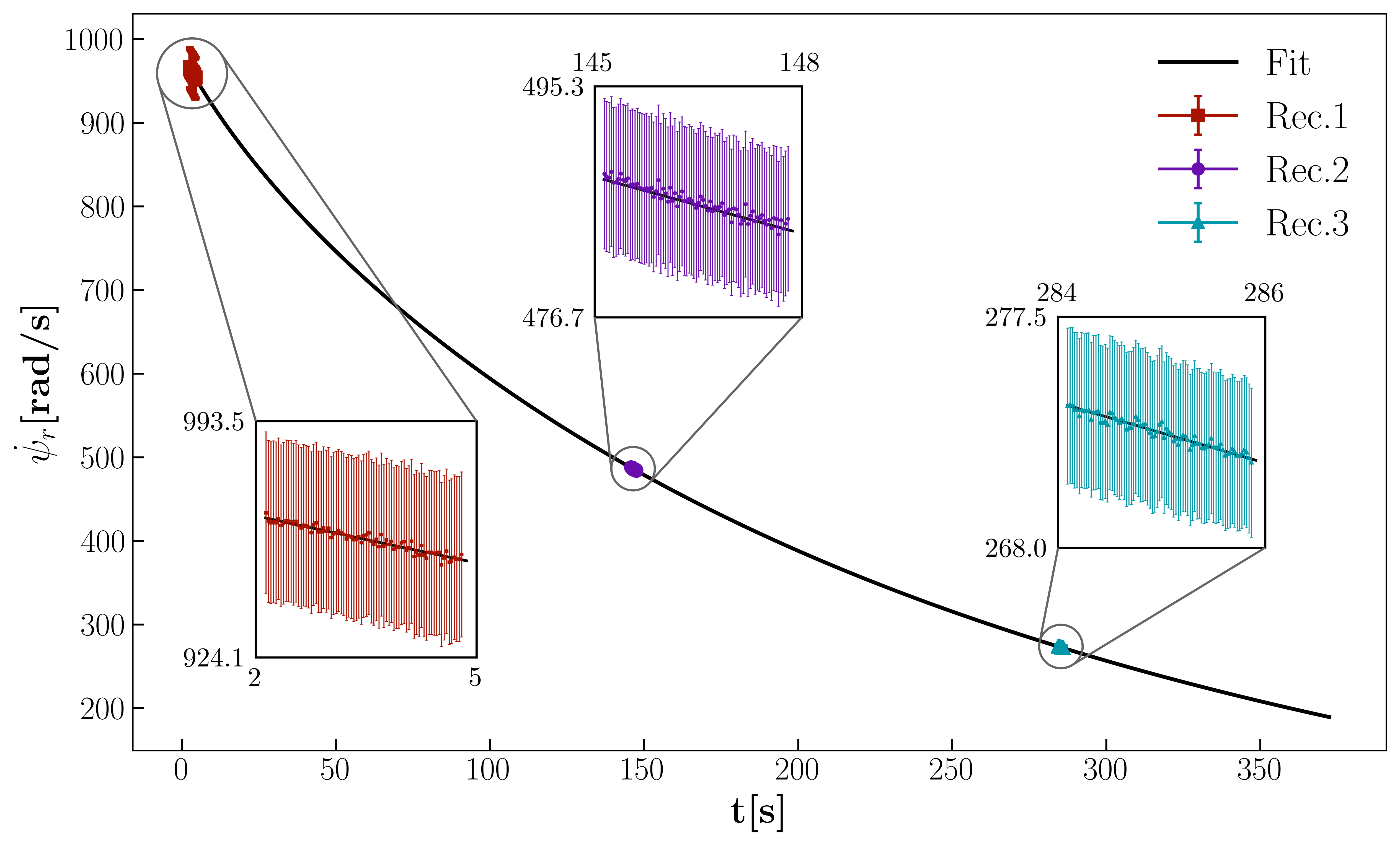}
    \caption{Case 2}
    \label{fig:second_gd}
  \end{subfigure}

	\caption{Experimental and generated data (GD) for the angular velocity of the rotor. Black lines show generated data and experimental data are seen with their uncertainties. There are three distinct experimental data for each case.}
  \label{fig:generated_data}
\end{figure}

The experimental setup can be seen in figure \ref{fig:exp_setup}.
The rotation of the casing is measured with an ELP Web camera, which takes images at 260 fps, while the rotation of the rotor is measured with a Sony ZV-1 camera taking images at 1000 fps for 3 seconds.
As a result, the measurement of the angular position of the rotor is not continuous, as seen in figure \ref{fig:generated_data}, which shows the angular velocities of the rotor.
The angular positions of the casing are also measured by the Sony ZV-1, and this data are used to synchronize the timing of the two measurements.

A Python code is used to track the locations of the center of the casing and the near edges of the colored pieces.
One colored paper piece is used to obtain the angular position of the casing.
The averaged position of the near corners from the center is taken as the angular position.
Conversely, three colored pieces are used to determine the angular position of the rotor to overcome the obstruction caused by the casing.
Unobserved points are ignored, and the remaining positions are used to determine the angular position of the rotor. 

The positions of the cameras, the rotation of the casing, and the angle of incoming light can affect the measurement of the angular positions, causing variations in each frame. 
By considering the base length of the colored piece $\delta$, the radius $\rho$, and the angle of the base with respect to the radius $\beta$, the uncertainty of each data point is calculated (see figure \ref{fig:gyroscope_p}).
To do this, the average radius and average base are calculated.
It is also worth noting that there is a tiny circular motion of the gyroscope alongside the main motion during the experiment, which contributes to the uncertainty of each data point (see the videos).

In an ideal case, the base length $\delta$ and radius $\rho$ are constant, and the angle between them $\beta$ is a right angle.
However, deviations occur for each data point due to the previously mentioned experimental factors.
These deviations from the ideal case are used to calculate uncertainties.
The first contribution to the uncertainty, $u_1$, is calculated by substituting the data base $\delta$ with the average base $\bar \delta$. 
The difference between the angular position of the data and the position generated using the average base equals $u_1$. 
Assuming a right angle and the average base length $\bar \delta$, another point is generated and used to calculate $u_2$ in a similar manner.
The third uncertainty, $u_r$, is calculated by comparing the average radius $\bar \rho$ with the radius of the data $\rho$ and utilizing the percentage difference.
These three sources of uncertainty are multiplied by specific constants to obtain the total uncertainty of each data point, using the following formula:
\begin{equation}
    \Delta \psi= \sqrt{(30 u_r)^2 +(150 u_1)^2+(150 u_2)^2} \label{unc}.
\end{equation}
This uncertainty is calculated for every tracked angular position of the near edges, and these results are then used to calculate the uncertainty of the final data.

The values for $u_r$, $u_1$, and $u_2$ are very small since every frame is evaluated independently.
However, during the experiment, the previously mentioned effects cause broader uncertainties, and there is no truly fixed point during the rotation of the gyroscope.
Considering this and other possible factors, the uncertainties are multiplied by the constants given in equation (\ref{unc}).

\begin{figure}[h!]
\centering
  \begin{subfigure}[b]{0.45\textwidth}
    \includegraphics[width=\textwidth]{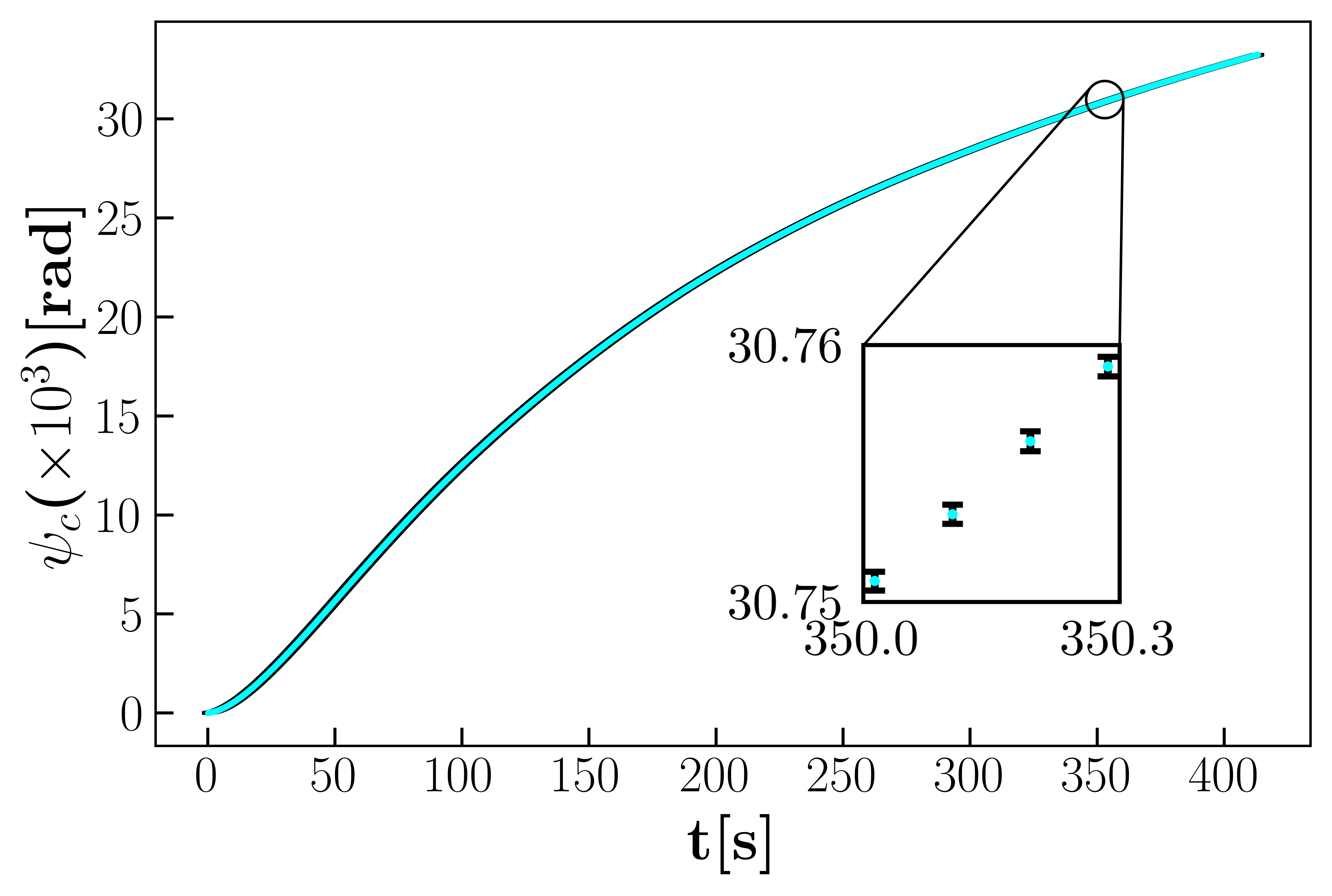}
    \caption{Case 1}
    \label{fig:first_ap}
  \end{subfigure}
  \hfill
  \begin{subfigure}[b]{0.45\textwidth}
    \includegraphics[width=\textwidth]{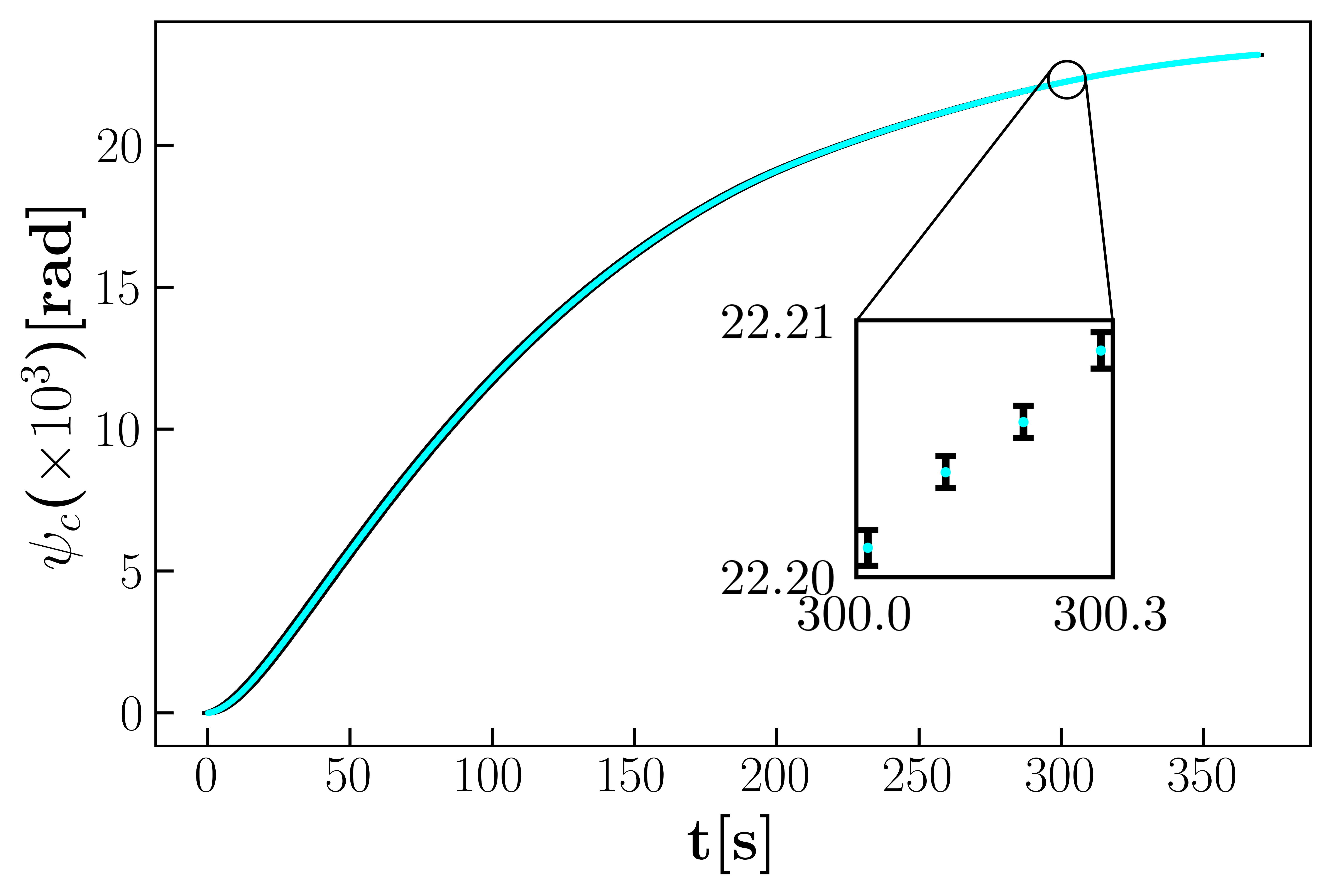}
    \caption{Case 2}
    \label{fig:second_ap}
  \end{subfigure}

  \caption{Angular positions of the casing for cases 1 and 2.}
  \label{fig:angular_positions}
\end{figure}

To smooth the angular positions, the data is median-averaged over 51 data points, with the first and last 50 data points omitted.
The graph of this averaged data for the casing can be seen in figure \ref{fig:angular_positions}.
The obtained angular positions are then used to calculate angular velocities, which are similarly median-averaged.
The rotational directions of the rotor and casing are clockwise (negative by convention); however, they are transformed to positive to avoid sign complications.
Since only one-dimensional angular rotation is considered, this does not pose any problems in this work.

As previously mentioned, due to equipment limitations, only fractions of the angular positions of the rotor are available.
Overcoming this issue is necessary to fit the data for the rotation of the casing.
Drawing on results from previous work \cite{Tanriverdi_dissipative}, two data sets for the angular velocity of the rotor's rotation were generated.
The experimental values and generated data are shown in figure \ref{fig:generated_data}.
From the figure, it can be seen that the generated data is consistent with the data and general structure obtained in \cite{Tanriverdi_dissipative}.
Using this generated data alongside the measured angular velocity of the casing, equations (\ref{dieqns}) are fitted to the data. 

\section{Fit tests}

A Python code is used to fit the parameters in equations (\ref{dieqns}) to the experimental results.
The code utilizes Radau to integrate the differential equations and least-squares to fit the parameters.
It fits the parameters to the data for both the angular velocities of the rotor and the casing.
Several scenarios are tested during the fitting procedure by setting different parameters to zero. 
For these tests, a smaller subset of the data is used to minimize computation time.

The moment of inertia of the casing is calculated as $I_c=2.06 \times 10^{-5} \; kg\cdot m^2$, and the rotor's is $I_r=5.54 \times 10^{-5} \; kg\cdot m^2$.
The dimensions of the fitted parameters are as follows: $[b_1]=[c]=ML^3/T^2$, $[a_1]=[b_2]=[d]=ML^3/T$, $[a_2]=[e]=ML^3$, $[f]=ML^3T^{g-2}$, and $g$ is dimensionless.
Although these parameters have dimensions, they are expressed without their units in the following sections.
It should be understood that they carry the mentioned dimensions in the SI unit system.

\begin{table}[h!]
        \centering
            \tiny
 \begin{center}
\begin{tabular}{|c|c||c|c||c||c|}
\hline
$a_1(10^{-11})$ & $a_2(10^{-11})$ & $b_1(10^{-6})$ & $b_2(10^{-7})$ &  $d(10^{-7})$   & $R^2$  \\
\hline
0.0432          & 11.0            & 7.94           & 7.56           & 2.34            & 0.993  \\
\hline
 0              & 11.0            & 7.93           & 7.56           & 2.34            & 0.993  \\
\hline
4080            & 0               & 0.00265        & 9.00           & 2.52            & 0.992  \\
\hline
\rowcolor{gray!25}
0.0317          & 10.2            & 0              & 8.62           & 2.39            & 0.989  \\
\hline
\rowcolor{gray!50}
13000           & 11.7            & 21.0           &    0           & 0.628            & 0.429  \\
\hline
\end{tabular}
\end{center}
        \normalsize
        \caption{Fitted parameters ($a_1, a_2, b_1, b_2, d$) and performance metric $R^2$ for casing data case 1 when $c=e=f=g=0$. Fit results satisfying $R^2\ge 0.99$ have a white background, $0.99> R^2\ge0.90$ have light gray, and $0.90> R^2\ge0.0$ have gray.}
        \label{tab:fit_results1_ab_c1}
\end{table}

\begin{table}[h!]
        \centering
            \tiny
 \begin{center}
\begin{tabular}{|c|c||c|c||c||c|}
\hline
$a_1(10^{-11})$ & $a_2(10^{-11})$ & $b_1(10^{-6})$ & $b_2(10^{-7})$ &  $d(10^{-7})$   & $R^2$  \\
\hline
\rowcolor{gray!25}
27.5            & 12.5            & 41.9            & 7.15           & 2.11            & 0.978  \\
\hline
\rowcolor{gray!25}
 0              & 12.7            & 42.2            & 7.07           & 2.10            & 0.978  \\
\hline
\rowcolor{gray!25}
6820            & 0               & 33.2            & 8.93           & 2.22            & 0.974  \\
\hline
\rowcolor{gray!50}
0.144           & 11.0            & 0              &13.6            & 2.22            & 0.849  \\
\hline
\rowcolor{gray!50}
12400           & 13.9            & 26.2            &    0           & 0.554           & 0.485  \\
\hline
\end{tabular}
\end{center}
        \normalsize
        \caption{Fitted parameters ($a_1, a_2, b_1, b_2, d$) and performance metric $R^2$ for casing data case 2 when $c=e=f=g=0$. Fit results satisfying $0.99> R^2\ge0.90$ have a light gray background, and $0.90> R^2\ge0.0$ have gray.}
        \label{tab:fit_results1_ab_c2}
\end{table}

Firstly, by choosing only the linear term for the interaction between the rotor and the casing, the effects of $a_1$, $a_2$, $b_1$, and $b_2$ are tested.
The results of these tests are presented in tables \ref{tab:fit_results1_ab_c1} and \ref{tab:fit_results1_ab_c2}. 
The tests show that $a_1$ does not significantly affect the fit quality and can be ignored.
$a_2$, $b_1$, and $b_2$ do affect the fit quality, so they are retained to test the friction models between the rotor and the casing.

\begin{table}[h!]
        \centering
            \tiny
 \begin{center}
\begin{tabular}{|c|c|c||c|c|c|c|c||c|}
\hline
$a_2(10^{-11})$ & $b_1(10^{-6})$ & $b_2(10^{-7})$ & $c(10^{-7})$ & $d(10^{-7})$ & $e(10^{-11})$ & $f(10^{-7})$ & $g$     & $R^2$  \\
\hline
9.31            & 5.56           & 8.21           & 48.2         & 1.97         & 7.65          & 0.406        & 0.587   & 0.996  \\
\hline
8.08            & 4.40           & 8.62           &   0          & 1.99         & 6.40          & 1.08         & 0.769   & 0.996  \\
\hline
9.21            & 5.01           & 8.29           & 28.2         &   0          & 6.58          & 1.68         & 1.04    & 0.996  \\
\hline
9.51            & 7.39           & 7.96           & 108          & 0.644        &  0            & 0.231        & 1.31    & 0.997  \\
\hline
11.0            & 9.60           & 7.45           & 71.8         & 1.82         & 7.60          &   0          & 0       & 0.996  \\
\hline
8.26            & 4.26           & 8.54           &   0          &    0         & 1.92          & 1.76         & 1.06    & 0.996  \\
\hline
9.06            & 5.13           & 8.29           &   0          & 2.01         &  0            & 0.0792       & 1.30    & 0.996  \\
\hline
8.25            & 4.50           & 8.58           &   0          & 2.32         & 5.14          & 0            &  0      & 0.996  \\
\hline
9.77            & 8.14           & 7.82           & 129          &  0           &  0            & 4.67         & 1.25    & 0.997  \\
\hline
\rowcolor{gray!25}
10.1            & 7.83           & 7.92           & 327          &  0           &  36.6         & 0            & 0       & 0.982  \\
\hline
11.0            & 7.93           & 7.56           &  0.00244     & 2.34         &  0            & 0            &  0      & 0.993  \\
\hline
8.92            & 4.86           & 8.35           &   0          &  0           &  0            & 1.76         & 1.06    & 0.996  \\
\hline
\rowcolor{gray!50}
4.46            & 15.3           & 7.69           &   0          & 0            &  77.7         & 0            &  0      & 0.578  \\
\hline
11.0            & 7.93           & 7.56           &   0          & 2.34         &  0            & 0            & 0       & 0.993  \\
\hline
\rowcolor{gray!75}
31.9            & $\approx 0$    & 4.84           &   347        & 0            &  0            & 0            & 0       &-0.285  \\
\hline
\end{tabular}
\end{center}
        \normalsize
    \caption{Fitted parameters ($a_2, b_1, b_2, c, d, e, f, g$) and performance metric $R^2$ for casing data case 1 when $a_1=0$. Fit results satisfying $R^2\ge 0.99$ have a white background, $0.99> R^2\ge0.90$ have light gray, $0.90> R^2\ge0.0$ have gray, and $0.0> R^2$ have dark gray. Where indicated as $\approx 0$, numbers are smaller than $1\times10^{-4}$.}
        \label{tab:fit_results_cdefg_c1}
\end{table}

\begin{table}[h!]
        \centering
            \tiny
 \begin{center}
\begin{tabular}{|c|c|c||c|c|c|c|c||c|}
\hline
$a_2(10^{-11})$ & $b_1(10^{-6})$ & $b_2(10^{-7})$ & $c(10^{-7})$ & $d(10^{-7})$ & $e(10^{-11})$ & $f(10^{-7})$ & $g$     & $R^2$  \\
\hline
7.84            & 41.6           & 9.00           & 102          & 1.35         & 16.0          & 2.22         & 0.453   & 0.995  \\
\hline
9.44            & 38.3           & 8.86           &   0          & 1.97         & 6.98          & 0.884        & 0.236   & 0.995  \\
\hline
13.1            & 43.8           & 6.63           & 103          &   0          & 10.8          & 1.62         & 0.962   & 0.991  \\
\hline
9.48            & 41.4           & 8.44           & 117          & 0.0256       &  0            & 0.321        & 1.30    & 0.996  \\
\hline
10.4            & 44.8           & 7.57           & 240          & 0.721        & 20.1          &   0          & 0       & 0.995  \\
\hline
\rowcolor{gray!25}
11.3            & 27.0           & 9.36           &   0          &    0         & 8.00          & 3.14         & 0.911   & 0.973  \\
\hline
8.41            & 37.5            & 9.33           &   0          & 0.916        &  0            & 0.223        & 1.31    & 0.996  \\
\hline
9.42            & 38.6           & 8.80           &   0          & 1.95         & 7.47          & 0            &  0      & 0.995  \\
\hline
10.6            & 44.0           & 7.62           & 192          &  0           &  0            & 0.153        & 1.40    & 0.995  \\
\hline
\rowcolor{gray!25}
12.0            & 18.6           & 10.0           & 215          &  0           &  33.4         & 0            & 0       & 0.972  \\
\hline
\rowcolor{gray!25}
13.1            & 43.8           & 6.69           &  1.20        & 2.07         &  0            & 0            &  0      & 0.976  \\
\hline
\rowcolor{gray!25}
11.6            & 42.0           & 7.51           &   0          &  0           &  0            & 1.62         & 1.05    & 0.986  \\
\hline
\rowcolor{gray!25}
17.8            & 21.1           & 7.06           &   0          & 0            &  35.9         & 0            &  0      & 0.948  \\
\hline
\rowcolor{gray!25}
12.7            & 42.2           & 7.07           &   0          & 2.10         &  0            & 0            & 0       & 0.978  \\
\hline
\rowcolor{gray!75}
28.9            & $\approx 0$    & 8.50           &   439        & 0            &  0            & 0            & 0       &-0.153  \\
\hline
 \end{tabular}
\end{center}
        \normalsize
        \caption{Fitted parameters ($a_2, b_1, b_2, c, d, e, f, g$) and performance metric $R^2$ for casing data case 2 when $a_1=0$. Fit results satisfying $R^2\ge 0.99$ have a white background, $0.99> R^2\ge0.90$ have light gray, and $0.0> R^2$ have dark gray. Where indicated as $\approx 0$, numbers are smaller than $1\times10^{-4}$.}
    \label{tab:fit_results_cdefg_c2}
\end{table}

For the next step, terms related to the interaction between the rotor and casing are studied.
The results of several scenarios are shown in tables \ref{tab:fit_results_cdefg_c1} and \ref{tab:fit_results_cdefg_c2}.
Because $g$ has no effect when $f$ equals zero, parameter $g$ is primarily evaluated alongside $f$.
According to the results, removing one of the four terms yields roughly the same fit quality in both cases.
If only $d$ or $f$ is non-zero (with the other three parameters set to zero), the fit quality remains reasonable.
Conversely, if only $e$ is non-zero, the fit quality diminishes compared to cases using only $d$ or $f$.
If only $c$ is non-zero, the fit results become illogical.
Thus, $c$ and $e$ do not significantly affect the fit quality, whereas $d$ and $f$ are the most effective terms.
The average value of $g$ is close to 1, however, the deviations from 1 for different special cases is substantial.
Then, by including the effect of parameter $d$, it is clear that the linear term is the one of the most effective terms, but it is not possible to say that fractional term, $f$, can be included in linear term.
From these result, it is possible to say that acceptable fit results with the fewest parameters can be achieved by including $a_2, b_1, b_2, d, f,$ and $g$.

\section{Fit}

The datasets for the angular velocities of the casing in cases 1 and 2 possess distinct properties:
Initial, maximum, and minimum values differ, which is advantageous for parameter determination. 
They also exhibit contrasting features such as peak structure and tail slope.
Additionally, unexpected spikes and troughs occur, the reasons for which are unclear.
One potential cause is the tiny circular motion of the casing running concurrently with the studied effect.
These minute rotations might stem from an asymmetry in the casing (i.e., $I_{xc}\ne I_{yc}$).

Based on the test results, a reduced model utilizing parameters $a_2$, $b_1$, $b_2$, $d$, $f$, $g$ is considered and compared with the generalized model including all terms considered in equation (\ref{dieqns}).
Fit results for the reduced model are presented in tables \ref{tab:fit_results_f1a} and \ref{tab:fit_metrics_fm_s}, while those for the generalized model appear in tables \ref{tab:fit_results_fp} and \ref{tab:fit_metrics_fm}.

Two approaches are applied: separate fits for each case, and a simultaneous joint fit.
During generalized-model fitting, parameter intervals are restricted to prevent large oscillations.
The ranges for $a_2$, $b_2$, and $g$ are set within $\pm 20\%$ of their average values derived from the reduced model fit.
For $d$ and $f$, a $50\%$ variance is allowed.
The ranges for parameters unused in the reduced model are kept wider, except for $g$, which is strictly constrained to $[0.3:1.4]$.

\begin{table}[h!]
        \centering
        \scriptsize
        \hspace{-2.0cm}
        \begin{center}
                \begin{tabular}{ |c |c|c|c|c|c|c|c|c| }
                        \hline
                &$a_2(10^{-11})$         & $b_1(10^{-6})$            & $b_2(10^{-7})$          & $d(10^{-7})$             & $f(10^{-7})$             & $g$          \\
                        \hline
                        \hline
            Case 1  &$9.97$                  &  $2.88$                   & $8.28$                  &  $1.57$                  &  $0.126$                 &  $1.32    $       \\
                        \hline
            Case 2  &$9.71$                  &  $42.2 $                  & $8.18$                  &  $1.12$                  &  $0.192$                 & $1.30$            \\
                        \hline
                                    \hline
            Case 1  &\multirow{2}{*}{$10.6$} &  $4.28 $                  & $8.19$                   & \multirow{2}{*}{$0.538$}& \multirow{2}{*}{$1.10 $} & \multirow{2}{*}{$1.08 $} \\
                        \cline{1-1} \cline{3-4}
            Case 2  &                        &  $44.0$                   & $7.54$                   &                         &                          &                      \\
                        \hline

                \end{tabular}
        \end{center}
        \normalsize
    \caption{Fitted parameters for the reduced model for case 1 and case 2 with full data.}
        \label{tab:fit_results_f1a}
\end{table}

\begin{table}[h!]
        \centering
        \scriptsize
        \hspace{-2.0cm}
        \begin{center}
                \begin{tabular}{ |c |c|c|c|c|}
                        \hline
                                                &         &$R^2$ for $\dot \psi_r$ & $R^2$ for $\dot \psi_c$   & Reduced $\chi^2$ for $\dot \psi_c$  \\
                        \hline
                        \hline
                        Fit 1                   & Case 1  &$1.000$                 &  $0.9968$                 & $2.520$                              \\
                        \hline
                        Fit 2                   & Case 2  &$0.9999$                &  $0.9934$                 & $9.781$                   \\
                        \hline
                        \hline
                        \multirow{2}{*}{Fit 3}  & Case 1  &  $0.9995$              &   $0.9861$                &  \multirow{2}{*}{$13.95$}    \\
                        \cline{2-4}
                                                & Case 2  &   $0.9996$             &    $0.9860$               &                                                      \\
                        \hline
                \end{tabular}
        \end{center}
        \normalsize
        \caption{Fit metrics for the reduced model.}
        \label{tab:fit_metrics_fm_s}
\end{table}

\begin{table}[h!]
        \centering
        \scriptsize
        \hspace{-2.0cm}
        \begin{center}
                \begin{tabular}{ |c |c|c|c|c|c|c|c|c|c|}
                        \hline
    &$a_1(10^{-11})$         &$a_2(10^{-11})$         & $b_1(10^{-6})$  & $b_2(10^{-7})$  & $c(10^{-7})$            & $d(10^{-7})$            &$e(10^{-11})$             & $f(10^{-7})$            & $g$          \\
                        \hline
                        \hline
Case 1  &$4020$                  &$7.96$                  &  $0.477$        & $7.16$          & $19.4$                  & $0.968$                 & $8.01 $                  & $0.162$                 &  $1.24 $      \\
                        \hline
Case 2  &$3830$                  &$8.02$                  &  $34.5 $        & $7.27$          & $154 $                  & $0.686$                 & $16.8 $                  & $0.0795$                & $1.05$        \\
                        \hline
                        \hline
Case 1  &\multirow{2}{*}{$3110$} &\multirow{2}{*}{$8.72$} &  $0.553$        & $7.59$          & \multirow{2}{*}{$89.5 $}& \multirow{2}{*}{$1.11 $}&  \multirow{2}{*}{$11.0$} & \multirow{2}{*}{$0.122$}&  \multirow{2}{*}{$1.09$} \\
                        \cline{1-1} \cline{4-5}
Case 2  &                        &                        &  $35.8$         & $7.24$          &                         &                         &                          &                         &  \\
                        \hline
                \end{tabular}
        \end{center}
        \normalsize
        \caption{Fitted parameters for the generalized model for case 1 and case 2 with full data.}
        \label{tab:fit_results_fp}
\end{table}

\begin{table}[h!]
        \centering
        \scriptsize
        \hspace{-2.0cm}
        \begin{center}
                \begin{tabular}{ |c |c|c|c|c|}
                        \hline
                                                &         &$R^2$ for $\dot \psi_r$ & $R^2$ for $\dot \psi_c$   & Reduced $\chi^2$ for $\dot \psi_c$  \\
                        \hline
                        \hline
                        Fit 1                   & Case 1  &$1.000$                 &  $0.9978$                 & $1.777$                              \\
                        \hline
                        Fit 2                   & Case 2  &$1.000 $                &  $0.9972$                 & $3.942$                   \\
                        \hline
                        \hline
                        \multirow{2}{*}{Fit 3}  & Case 1  &  $0.9995$              &   $0.9816$                &  \multirow{2}{*}{$11.43$}    \\
                        \cline{2-4}
                                                & Case 2  &   $0.9997$             &    $0.9923$               &                                                      \\
                        \hline
                \end{tabular}
        \end{center}
        \normalsize
        \caption{Fit metrics for full data utilizing the generalized model.}
        \label{tab:fit_metrics_fm}
\end{table}

Next, the fitted parameters for both the reduced and generalized models are evaluated.
Multiplying a parameter by the appropriate power of corresponding angular velocity yields the corresponding torque (see equations (\ref{dieqns}) and (\ref{frictiont})).
Observing the variation in angular velocities and the magnitude of the fitted parameters, each torque term is determined to lie within the order of $10^{-5}-10^{-7}$.
The least effective term is $c$, at the order of $10^{-7}$; the remaining terms vary slightly, largely dependent on the angular velocities.
Consequently, this comparison merely establishes $c$ as the least influential parameter.

The robustness of the parameters can also serve as an evaluation metric.
For the reduced model: 
1) $a_2$ and $b_2$ are the most robust parameters. Thus, it can be inferred that $b_2$ is independent of the contact surface.
2) The variation of $b_1$ between cases relates to the contact surface, while its shift from individual to joint fits remains minor. Then, it is possible to say friction between the touchpoint and the ground is mainly Coulomb friction.
3) Parameters $d$, $f$, and $g$ exhibit substantial variation between cases and during the joint fit, indicating that their determination is not optimally precise.

For the generalized model: 
1) Similar to the reduced model, $a_2$ and $b_2$ remain highly robust. 
2) $b_1$ changes depending on the surface and proves more robust than in the reduced model. 
3) Variations in $a_1$, $d$, and $g$ fall within acceptable limits for this study, though they are still noticeable. 
4) Large oscillations occur in $c$, $e$, and $f$, suggesting their values can only be approximated. 

Table \ref{tab:fit_metrics_fm_s} displays the fit metrics for the reduced model. The $R^2$ values are nearly 1 for the rotor and are acceptable for the casing.
As expected, the joint fit performs slightly worse than the separate fits.
Because the rotor data is generated and lacks uncertainty bounds, reduced $\chi^2$ values can only be discussed for the casing; these are acceptable for case 1 but worsen for case 2 and the joint fit.
The generalized model improves fit metrics due to its additional parameters, but this negatively impacts parameter robustness.

While both models yield strong metrics, evaluating parameter robustness across fits favors the reduced model.
In the generalized model, parameters defining the casing-rotor interaction fluctuate substantially between cases.
This implies the generalized model may overfit these two cases, with some parameters compromising the robustness of others.
Such instability hinders predictive capabilities; therefore, the reduced model is preferred in this regard.

\begin{figure}[h!]
\centering
  \begin{subfigure}[b]{0.45\textwidth}
    \includegraphics[width=\textwidth]{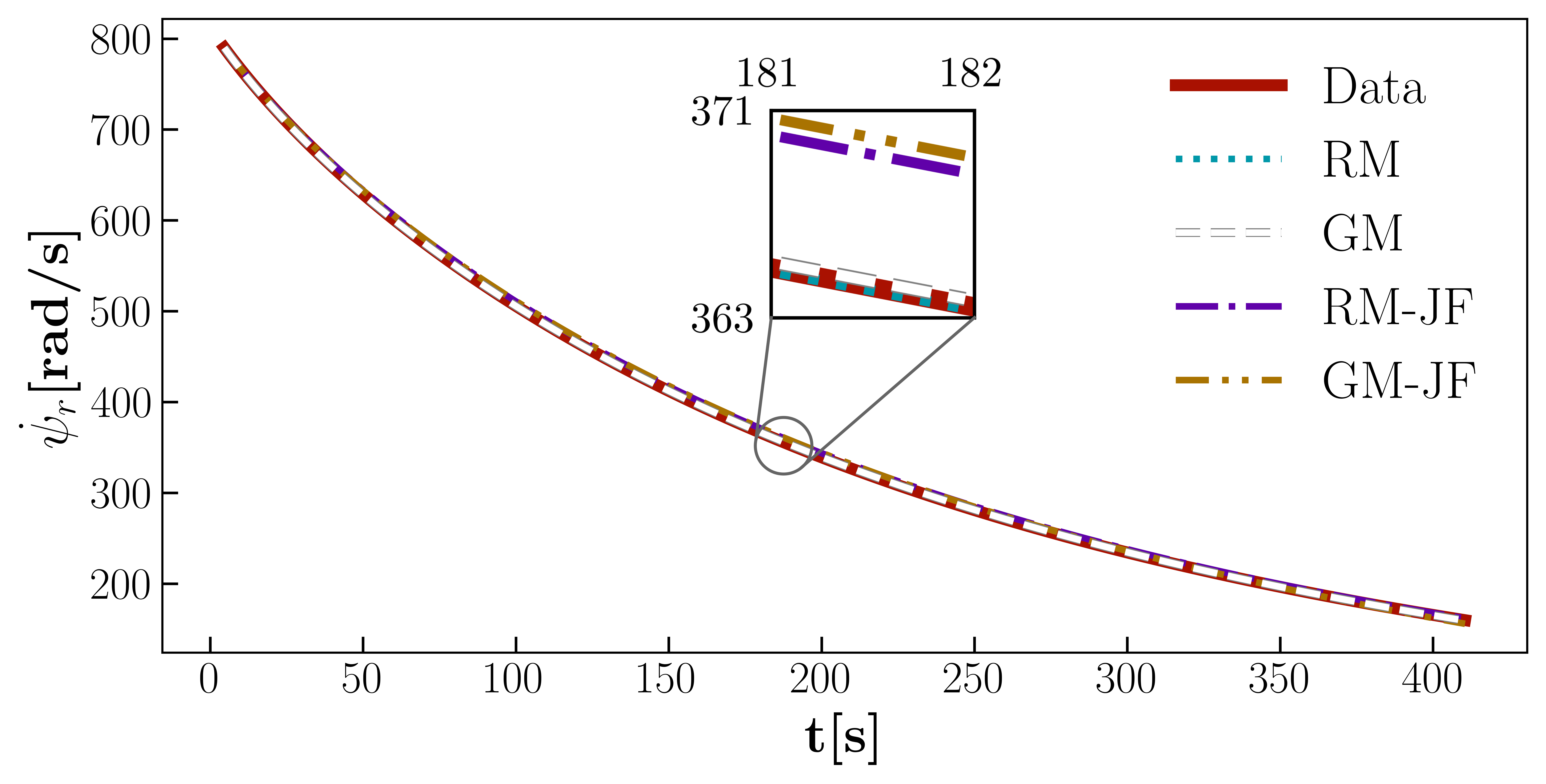}
    \caption{Case 1}
    \label{fig:first_avr}
  \end{subfigure}
  \hfill
  \begin{subfigure}[b]{0.45\textwidth}
    \includegraphics[width=\textwidth]{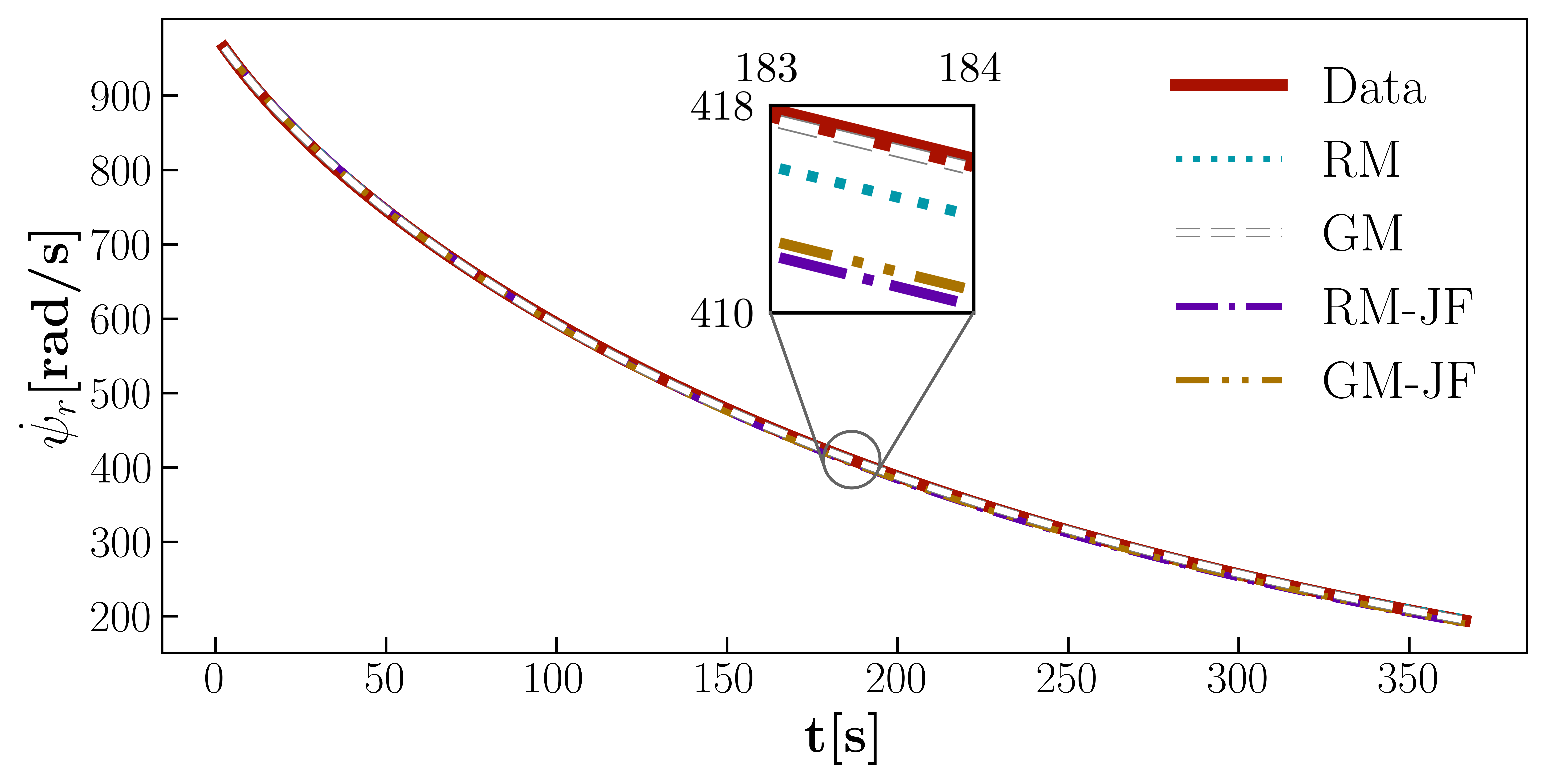}
    \caption{Case 2}
    \label{fig:second_avr}
  \end{subfigure}

	\caption{Generated data and fits for the rotor's angular velocity. Continuous (red) lines show the data generated according to experimental results. There are four different fit results: Dotted (turquoise) lines show reduced model (RM), dashed (white) lines shows generalized model (GM), dotted-dashed (purple) lines show reduced model obtained from joint fit (RM-JF), and double dotted-dashed (brown) lines show generalized model obtained from joint fit (GM-JF).}
  \label{fig:ang_vel_rot}
\end{figure}

\begin{figure}[h!]
\centering
  \begin{subfigure}[b]{0.45\textwidth}
    \includegraphics[width=\textwidth]{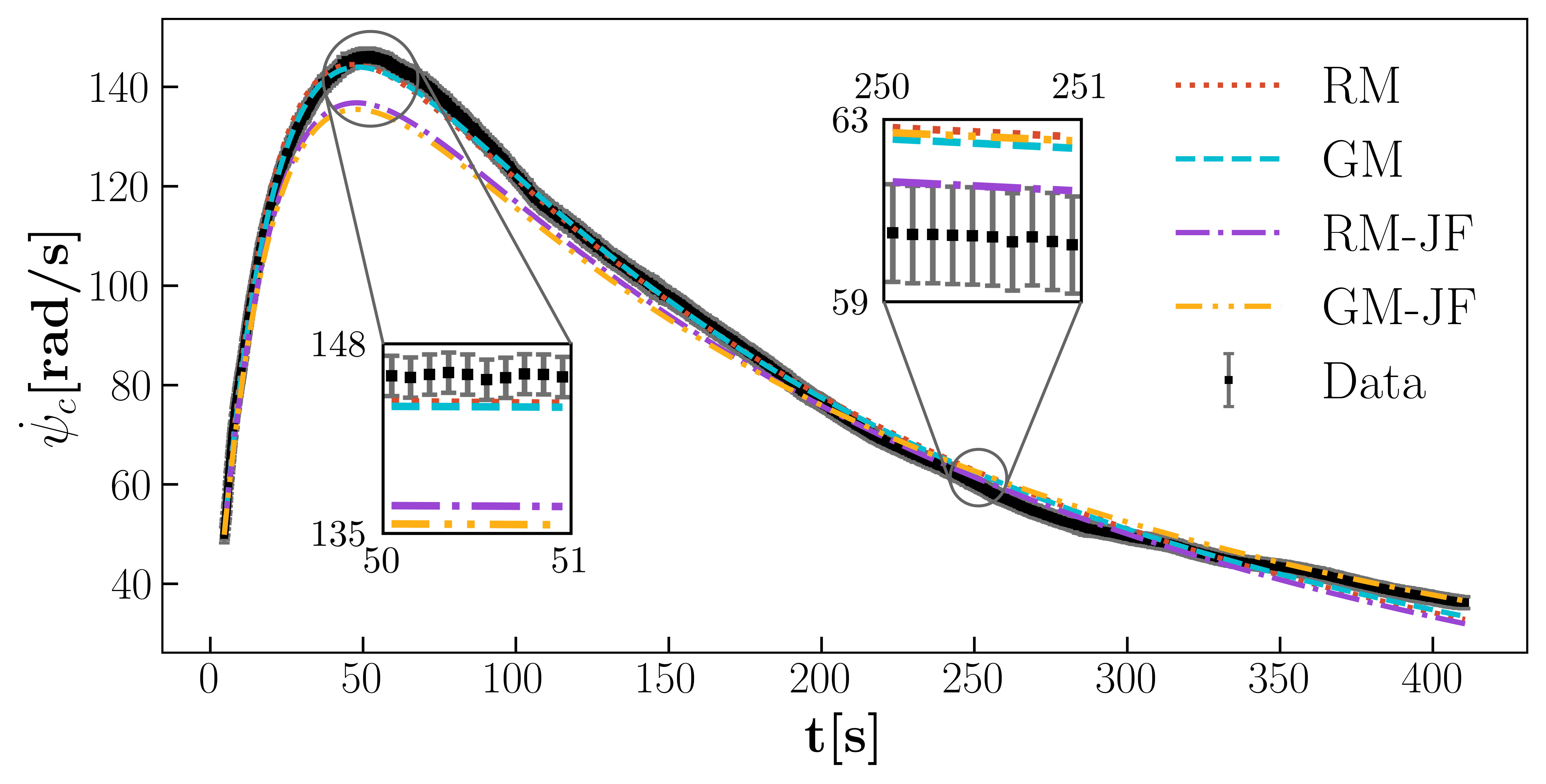}
    \caption{Case 1}
    \label{fig:first_avc}
  \end{subfigure}
  \hfill
  \begin{subfigure}[b]{0.45\textwidth}
    \includegraphics[width=\textwidth]{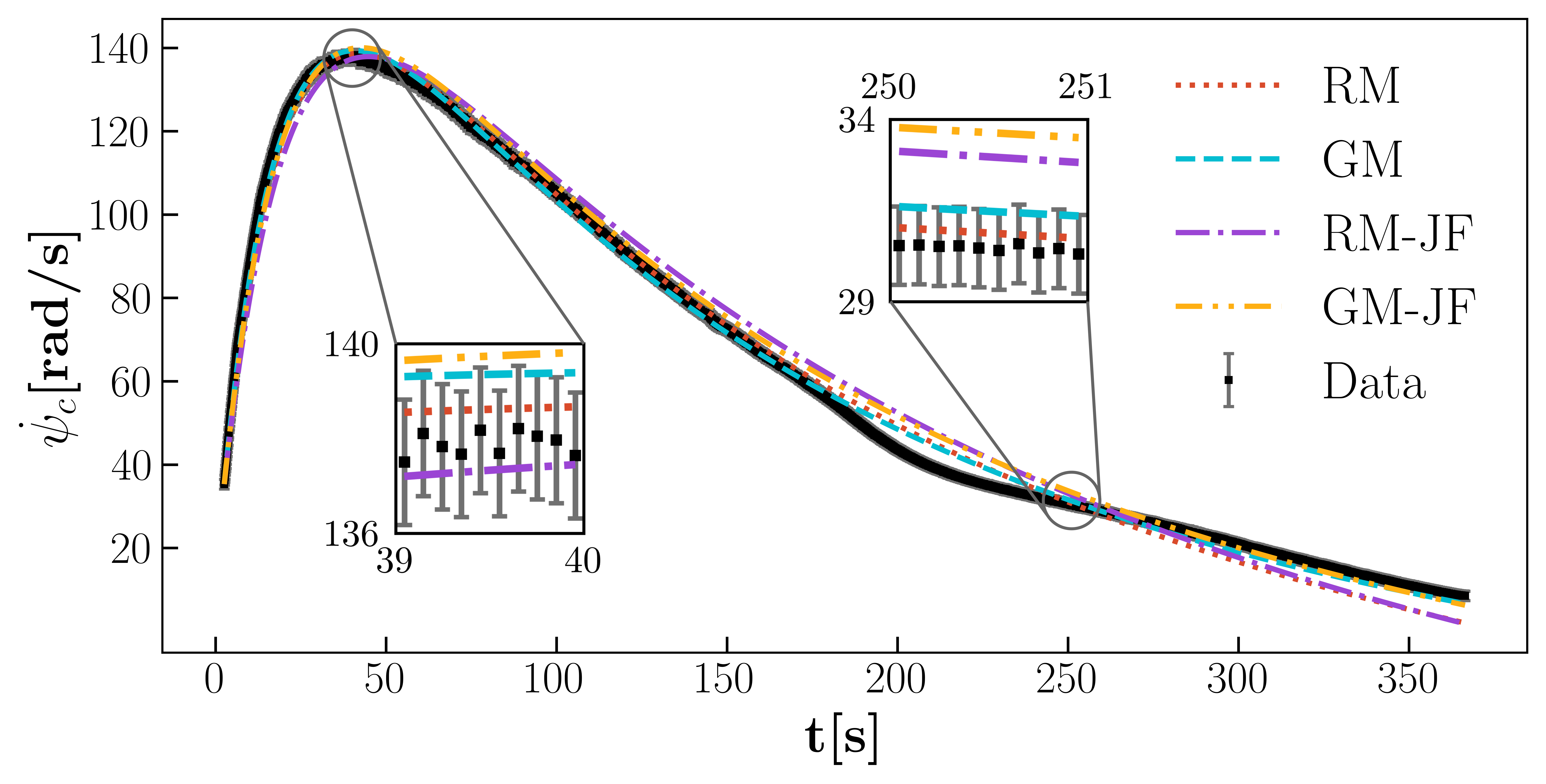}
    \caption{Case 2}
    \label{fig:second_avc}
  \end{subfigure}

	\caption{Angular velocities of the casing. Experimental data is shown by (black) dots together with their uncertainties. There are four different fit results: Dotted (red) lines show reduced model (RM), dashed (turquoise) lines shows generalized model (GM), dotted-dashed (purple) lines show reduced model obtained from joint fit (RM-JF), and double dotted-dashed (yellow) lines show generalized model obtained from joint fit (GM-JF).}
  \label{fig:ang_vel_casing}
\end{figure}

\begin{figure}[h!]
\centering
  \begin{subfigure}[b]{0.45\textwidth}
    \includegraphics[width=\textwidth]{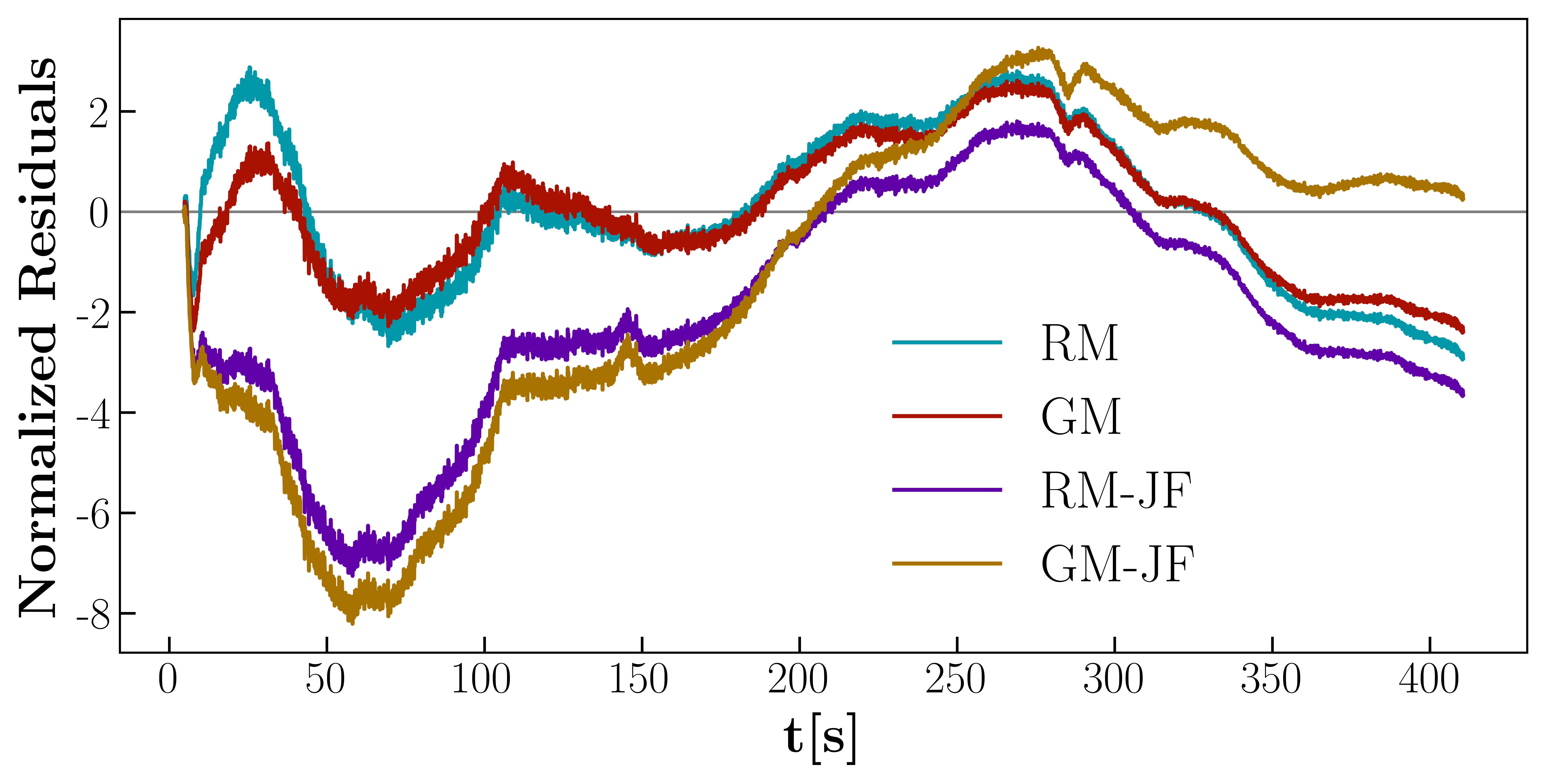}
    \caption{Case 1}
    \label{fig:first_res}
  \end{subfigure}
  \hfill
  \begin{subfigure}[b]{0.45\textwidth}
    \includegraphics[width=\textwidth]{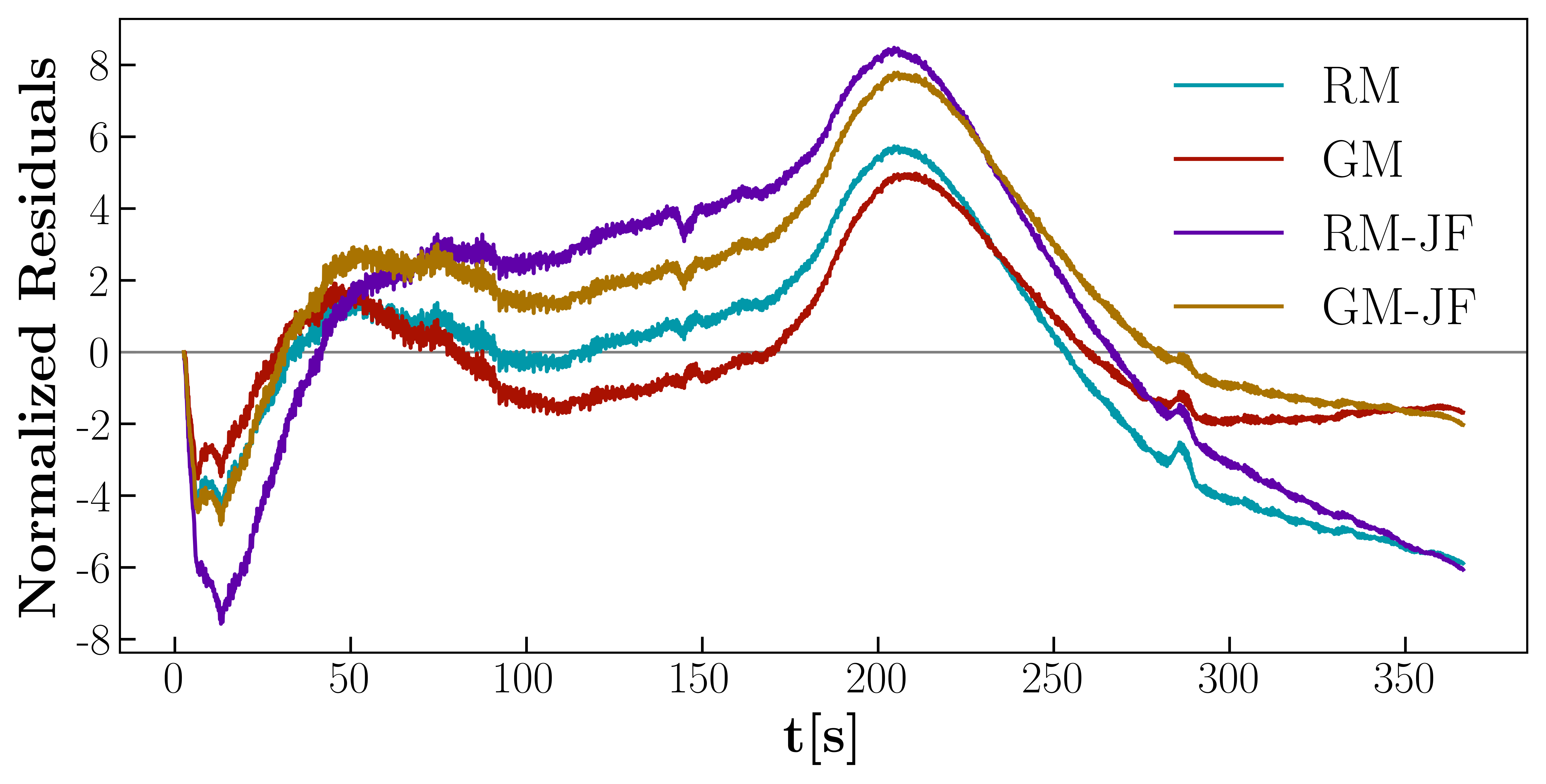}
    \caption{Case 2}
    \label{fig:second_res}
  \end{subfigure}

	\caption{Normalized fit residuals for the casing's angular velocity, $(\dot \psi_{c_{fit}}-\dot \psi_{c_{exp}})/\sigma$ where $\sigma$ represents the experimental uncertainty. Turquoise lines show reduced model (RM), red lines show generalized model (GM), purple lines show reduced model obtained from joint fit (RM-JF), brown lines show generalized model obtained from joint fit (GM-JF).}
  \label{fig:residuals}
\end{figure}

The resulting fit curves are plotted alongside the data in figures \ref{fig:ang_vel_rot} and \ref{fig:ang_vel_casing}. 
As shown in figure \ref{fig:ang_vel_rot}, the models demonstrate very good agreement with the generated data for the angular velocity of the rotor, $\dot \psi_r$. 
For the angular velocity of the casing, $\dot \psi_c$, however, slight deviations between the fits and the experimental data emerge, primarily near the peak and along the tail (figure \ref{fig:ang_vel_casing}). 
These discrepancies are more clearly illustrated in the normalized residual plots in figure \ref{fig:residuals}. 
Rather than scattering randomly around zero, the residuals exhibit structured patterns in these specific regions. 
According to the normalized residuals, the independent single fits using the generalized model yield the most statistically sound results. 
As expected, the systematic patterns in the residuals become more pronounced when employing the reduced model or joint fits. 
While some of these structural patterns correspond to unexpected, localized peaks and troughs observed in the raw data, overall, the generalized model provides a highly acceptable description of the kinematics of the system.

\section{Energy approach}

The problem can also be addressed energetically.
For a dissipative torque, it is possible to write
\begin{equation}
    \Delta K=-\int \tau \dot \psi dt, \label{energy}
\end{equation}
where $\Delta K$ is the change in rotational kinetic energy, $\tau$ is the dissipative torque, and $\psi$ defines the angle of rotation.
This equation dictates that the change in rotational kinetic energy must equal the energy dissipated.
Rewriting this as $K_f+\int \tau d \psi=K_i$, one expects $K_f+\int \tau d \psi$ to remain constant and equal to the initial kinetic energy $K_i$.

The analyzed models can be tested by checking the constancy of $K_f+\int \tau d \psi$.
Here, $K$ is the sum of the rotational kinetic energies of the rotor and the casing.
Thus, for this scenario: 
\begin{equation}
K_i=K_f+\int [-\tau_{cr}(\dot \psi_r-\dot \psi_c)+(-a_1 \dot \psi_r-a_2 \dot \psi_r^2)\dot \psi_r+(-b_1-b_2\dot \psi_c)\dot \psi_c]dt . \label{energy2}
\end{equation}

\begin{figure}[h!]
\centering
  \begin{subfigure}[b]{0.45\textwidth}
    \includegraphics[width=\textwidth]{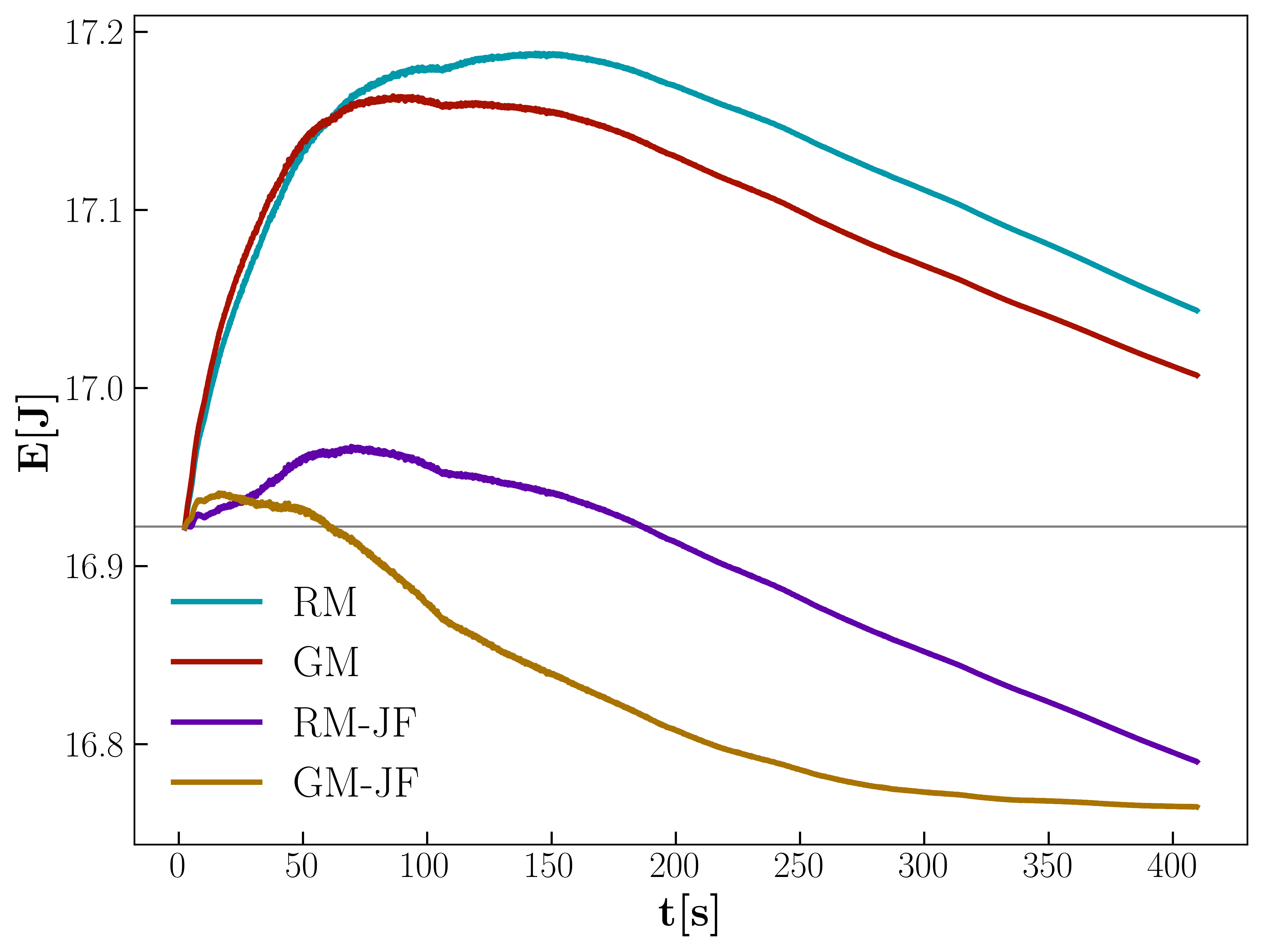}
    \caption{Case 1}
    \label{fig:first_en}
  \end{subfigure}
  \hfill
  \begin{subfigure}[b]{0.45\textwidth}
    \includegraphics[width=\textwidth]{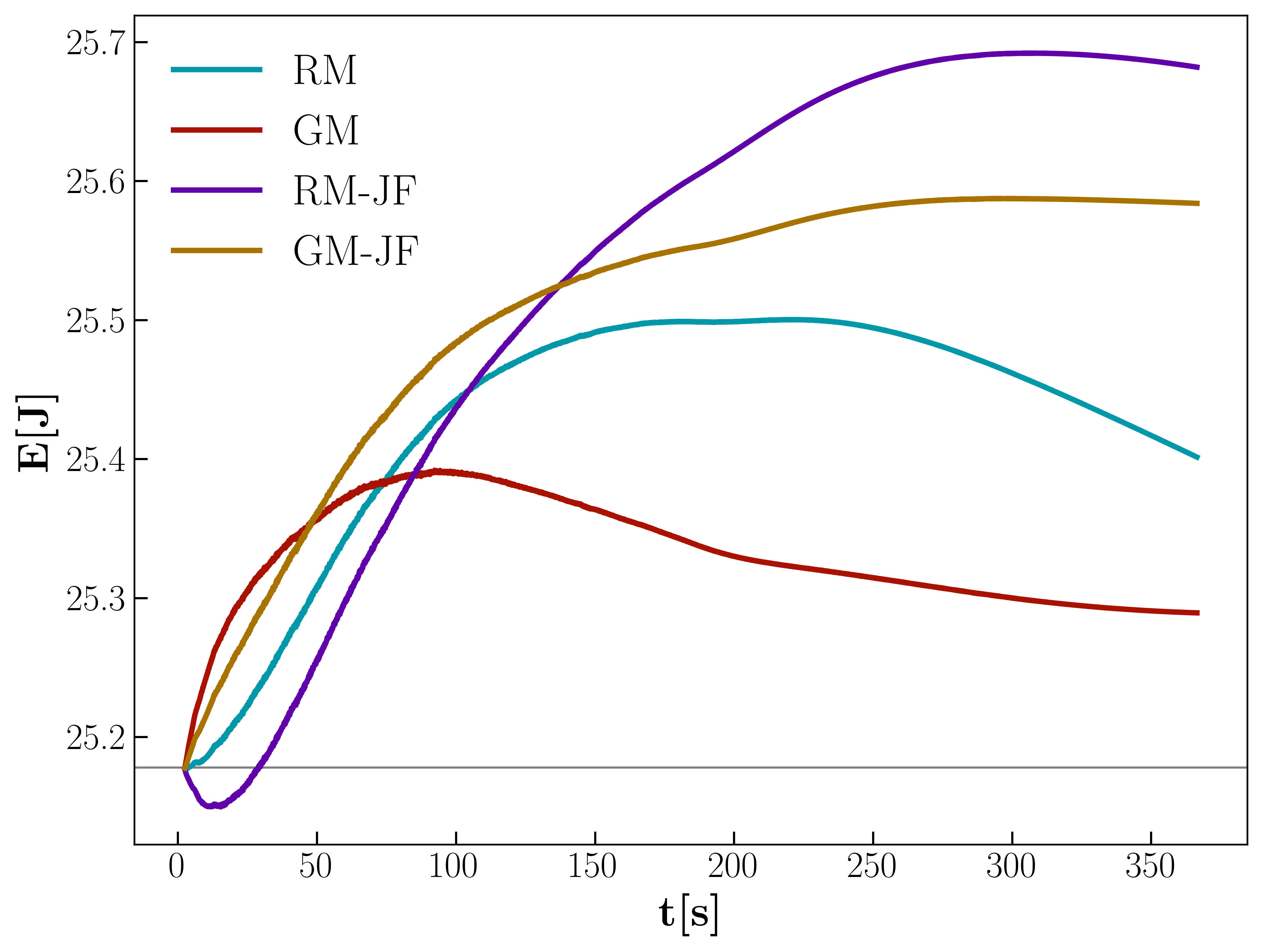}
    \caption{Case 2}
    \label{fig:second_en}
  \end{subfigure}

  \caption{Change in $K_f+\int \tau d \psi$ across different cases. Turquoise lines show reduced model (RM), red lines show generalized model (GM), purple lines show reduced model obtained from joint fit (RM-JF), brown lines show generalized model obtained from joint fit (GM-JF).}
    \label{fig:energychange}
\end{figure}

Using the parameters derived from the models, the right-hand side of equation (\ref{energy2}) can be calculated via the experimental angular velocities, providing an energetic consistency check for the fits. 
Plots of the right-hand side of equation (\ref{energy2}) for these fits appear in figure \ref{fig:energychange}.
For case 1, figure \ref{fig:first_en} indicates that the reduced model with a joint fit performs best energetically, whereas the reduced model with a single fit performs the worst. 
Conversely, for case 2, figure \ref{fig:second_en} shows that the generalized model with a single fit yields the optimal result, while the reduced model with a joint fit performs the worst. 
Ultimately, the energy approach does not uniformly favor a single modeling strategy; an approach yielding optimal results in one case can prove suboptimal in the other. 
However, the discrepancies from the initial kinetic energy are only around 2\%, indicating that all tested models remain largely consistent with energy considerations.

\section{Summary and Conclusion}

The frictional effects driving the rotation of a casing and rotor due to initial rotor spin were studied.
Various friction terms were modeled and tested against the experimental data.
While several models can capture the broad characteristics of this rotation, achieving high precision remains a challenge.
Distinct friction models often yield statistically similar results, making definitive model selection difficult without further study though the single case modeling strategy with generalized model gives most consistent results.

Two primary sources of uncertainty identified in this work are a potential asymmetry of the casing, which may induce the observed minute circular motions, and the temperature dependence of friction parameters.
The residuals in our fit analysis exhibit patterns suggesting either unmodeled physical effects or experimental artifacts related to these factors.
Resolving these uncertainties will require further experimentation with higher-precision equipment and thermal control.

By analyzing the quantity $K_f+\int \tau d \psi$, the consistency of the dissipative models is evaluated.
Variations in this quantity highlight discrepancies that standard fit statistics might miss.
To our knowledge, such an energy-conservation metric has not been previously utilized to discriminate between friction models in this context though similar methods are used to study friction at the ball bearing \cite{wu_he}.

Beside scientific curiosity, mastering this type of motion may aid the development of enhanced models for gyroscopically controlled satellites.
The average rotor angular velocities examined in this work are comparable to the operating speeds of CMGs in International Space Station ($691rad/s$) \cite{Gurrisi}.
As noted in the introduction, LuGre model is frequently used for modeling the friction.
The fit tests conducted herein include constant and velocity-dependent terms analogous to the LuGre model.
The findings suggest that LuGre-based models require updates to accurately capture higher-speed rotations (see table \ref{tab:fit_results_cdefg_c1} and \ref{tab:fit_results_cdefg_c2}).
Developing more accurate friction models for these high-velocity regimes can be crucial for predicting torque variances and extending the operational lifespan of CMGs.

It should be noted that the operational environment for satellites and space telescopes is vastly more complex than the system studied here, as they rotate in three-dimensional space.
Furthermore, the presence of a motor introduces additional rotating components.
Other potential complexities include temperature dependencies and spin-orbit interactions.
CMG aging also shifts friction parameters over time, with torsion being a primary catalyst for these changes.
Reducing CMG rotational speeds can minimize torsion and mitigate aging.
The methodologies investigated in this work can be adapted to refine the modeling of space telescope rotations governed by CMGs.
Instead of modeling friction, rotations as a result of CMG's rotation can be modeled similar to modeling the rotation of the casing in this work and the better results can be obtained.

\section{Acknowledgement}

During the preparation of this work, the authors utilized Gemini and ChatGPT to assist with language refinement, code structure, or literature finding. Specifically, Gemini (Google, Gemini plus) was used for language editing and proofreading, and ChatGPT (OpenAI, GPT-4 pro) was used for Python codes. All outputs were critically reviewed and validated by the authors, who take full responsibility for the final content.

The authors thank Musa Rajamov for introducing the problem.

\section{Appendix}

The videos can be viewed at the following links:

Case 1: 

Casing recording \url{https://youtu.be/OC0rItjdHyo},

1000 fps Record 1 \url{https://youtu.be/nIwzbinMMiE}

1000 fps Record 2 \url{https://youtu.be/ZEAqe2OQWgQ}

1000 fps Record 3 \url{https://youtu.be/0De0CrpnVEU}

Case 2: 

Casing recording \url{https://youtu.be/2g7ui-fXtUY}

1000 fps Record 1 \url{https://youtu.be/0UPXc_WhS_8}

1000 fps Record 2 \url{https://youtu.be/227SjOoGhgY}

1000 fps Record 3 \url{https://youtu.be/Z8hgOyxeCAQ}

\end{document}